\documentclass[aps,pre,twocolumn,floatfix,nofootinbib,showpacs,amsmath,amssymb]{revtex4}
\usepackage[english]{babel}
\usepackage{graphicx}
\usepackage{textcomp}
\usepackage{amsmath}
\def\sign{\mathop{\rm sgn}\nolimits}
\begin{document}
\title{Exactly solvable dynamics of the Eigen and
the Crow-Kimura models}
\author{David B. Saakian$^{1,2}$}
\author{Olga Rozanova $^3$}
\author{Andrei Akmetzhanov $^4$}
\email{saakian@yerphi.am} \affiliation{$^1$Yerevan Physics
Institute, Alikhanian Brothers St. 2, Yerevan 375036, Armenia}
\affiliation{$^2$Institute of Physics, Academia Sinica, Nankang,
Taipei 11529, Taiwan} \affiliation{$^3$Mathematics and Mechanics
faculty, Moscow State University, Main building, Moscow 119992
Russia} \affiliation{$^4$Institute for Problems in Mechanics,
Russian Academy of Sciences, Vernadsky Ave. 101-1, Moscow 115926,
Russia}
\date{\today}
\date{\today}
\begin{abstract}
We introduce a new way to study molecular evolution within
well-established Hamilton-Jacobi formalism, showing that for a broad
class of fitness landscapes it is possible to derive dynamics
analytically within the $1/N$-accuracy, where $N$ is genome length.
For smooth and monotonic fitness function this approach gives two
dynamical phases: smooth dynamics, and discontinuous dynamics. The
latter phase arises naturally with no explicite singular fitness
function, counterintuitively. The Hamilton-Jacobi method yields
straightforward analytical results for the models that utilize
fitness as a function of Hamming distance from a reference genome
sequence. We also show the way in which this method gives dynamical
phase structure for multi-peak fitness.
\end{abstract}
\pacs{87.10.+e, 87.15.Aa, 87.23.Kg, 02.50.-r} \maketitle
\section{\bf Introduction}
Genome dynamics is an important problem in population genetics [1-3]
and in molecular evolution [4-9]. Many authors investigated dynamics
of evolution [10-13]. The Crow-Kimura and the Eigen models are very
popular in evolution theory, describing quite well population
genetics, the RNA virus evolution, and artificial evolution of
molecules. The Crow-Kimura model describes evolutionary process
where mutation and selection are two parallel processes and
describes mutations during the life time. The Eigen model describes
the case where mutations occur during the birth of new viruses
(molecules) and is quite realistic for the RNA virus evolution.
While exact solution is known for a simple case of single-peak
fitness [14-16], there has been no success thus far in calculating
exact dynamics for a general fitness landscape. As in molecular
evolution, there are numerous attempts to solve this problem at
least approximately [10-13]. The fact is that evolution models are
very subtle mathematical objects and approximate solutions often
give misleading or inadequate results, especially in dynamics.
Finding exact dynamics for these two models is well-known to be
still an open issue. In this article we introduce Hamilton-Jacobi
equations (HJE) as a mean to resolve it. These equations have been
already applied in evolution theory to investigate population
genetics of virus evolution with a finite population \cite{ro03}. In
Ref.~\cite{ro03} HJE were applied and solved approximately for
linear fitness. Also, HJE had been utilized in Refs.~[18-19] to
derive exact steady-state solutions for evolution models with a
general fitness. In this work we show that it is possible to obtain
exact dynamical solutions of the Hamilton-Jacobi equations for the
models where fitness is defined in terms of the Hamming distance
from a reference (wild) sequence. The possibility of having
analytical solutions that give the dynamics in a closed form is an
important breakthrough in the theory of biological evolution. It
allows one the investigation of a plethora of evolutionary pathways
within one consistent formalism. By mapping evolution model to
Hamiltonian mechanics and looking at the corresponding potential, it
is possible to derive phase structure of the dynamics when exact
dynamics are unavailable by other means. We show here the way to
precisely calculate the movement of the maximum of the distribution
for the population originally localized at a fixed distance from a
reference sequence. This article is organized as follows. In Sec.~II
we review the known results for the Crow-Kimura model, analyze its
dynamics via HJE when population is initially localized at some
Hamming distance from a reference sequence, and investigate the case
when originally population is uniformly distributed across the
sequence space. In Sec.~III we solve the dynamics of the Eigen
model. Our results are discussed in Sec.~IV.
\section{\bf The Crow-Kimura model}
\subsection{Main known results}
The $2^N$ genome configuration sequences are defined as chains of
$N$ spins $s_n, 1\le n \le N$, that can take on only two values
$s_n=\pm 1$. The reference configuration has all spins $+1$. The
Hamming distance between a given configuration and the reference
configuration is $\sum_n(1-s_n)/2 = N(1-m)/2$, where $m$ is an
overlap. This model describes the dynamics of probability
distribution. We denote configuration $i$ by $S_i\equiv ({s_i^1,
\dots, s_i^N})$. The state of the system is specified by $2^N$
relative frequencies $P_i, 1 \le i \le 2^N$:
\begin{eqnarray}
\label{e1}
\frac{{dP}_i}{dt}=\sum_jA_{ij}P_j-P_i\sum_jP_jr_j ,\nonumber\\
A_{ij}=\delta_{ij}r_j+m_{i j}.
\end{eqnarray}
Here $m_{ij}$ is the rate of mutation from configuration $S_j$ to a
new configuration $S_i$, and $r_{i}$ is the fitness. Two
configuration states have a Hamming distance $d_{ij}=(N-\sum_k s_i^k
s_j^k)/2$, and $m_{ii}=-\gamma_0 N$. When $d_{ij}=1$ then
$m_{ij}=\gamma_0$ and $m_{ij}=0$ for $d_{ij}>1$ \cite{ba97}. For
index $i$, the set of values $1\le i \le 2^N$ is equivalent to the
collection of $N$ spins $s_k$. Identifying $f_0(s_1...s_N)\equiv
r_i$, we define the mean fitness $R$:
\begin{eqnarray}
\label{e2}R\equiv \sum_iP_ir_i.
\end{eqnarray}
The model defined here by Eq.(\ref{e1}) had been introduced in
Ref.~[3] to describe the Drosphilla's evolution in a multi-allele
model with simultaneously present mutation and selection processes.
Because this model describes genetics of diploid evolution in
infinite population the random drift is necessarily absent. The
diploid evolution model of Ref.[3] is described by an equation in
analogy with Eq.(\ref{e1}) except that $r_i$ are linear functions of
$p_i$. In the model of [4] our Eq.(\ref{e1})  describes an infinite
population asexual evolution when there are either many alleles in
one locus or many loci with two alleles in each. The selection and
mutation processes are decoupled in Eq.(\ref{e1}), i.e., our model
describes selection and mutation as parallel processes. This is
different to a well-known model introduced by Eigen
\cite{ei71,ei89}, where it is assumed that mutations originate as
replication errors on the occasion of reproduction events. Nowadays
the Eigen's model is widely applied to describe the virus evolution.
The model of Ref.~\cite{ba97} as well as the Eigen's model
\cite{ei71,ei89} have been suggested as molecular evolution models.
Both these ``connected mutation-selection" schemes of
Refs.~\cite{ei71,ei89} and "parallel","decoupled" scheme of
Ref.~\cite{ba97} are similar, giving similar pictures of evolution
with only a slight difference in dynamics (e.g., see Fig.~1 in
Ref.~\cite{sh04b}). The difference between the connected
multi-selection scheme and the parallel mutation-selection scheme of
this work becomes transparent when both models are treated by a
quantum Hamiltonian approach \cite{ba97,sh04a}: the parallel scheme
is described in terms of Hermitian Hamiltonian and the connected
scheme is described in terms of non-Hermitian Hamiltonian.

A value of $R$ in steady state ($dP_i/dt=0$) is the main target of
theoretical investigations. One can calculate $R$ as maximal
eigenvalue of a matrix $A_{ij}$ \cite{ei89,bw01}. Connection between
the Crow-Kimura model and quantum mechanics has been established in
Ref.~\cite{ba97}, where matrix $-A_{ij}$ has been identified with
the quantum Hamiltonian $H$ for $N$ interacting quantum spins. One
can calculate the maximal eigenvalue of the operator $-H$
\cite{bw01,sh04c} as
\begin{eqnarray}
\label{e3} R=\lim_{\beta\to\infty}\frac{\ln Tr \exp[-\beta
H]}{\beta},
\end{eqnarray}
where
\begin{eqnarray}
\label{e4}
-H=\gamma_0\sum_{k=1}^N(\sigma^x_k-1)+f_0(\sigma^z_1...\sigma^z_N),
\end{eqnarray}
where $\sigma^z_k$ and $\sigma^x_k$ are Pauli matrices acting on the
spin in the $k$th position \cite{sh04c}. We are interested in
symmetric-fitness case with $f_0(s_1...s_N)\equiv
Nf(\sum_{k=1}^Ns_k/N)$. For a symmetric fitness function and
permutation-symmetric initial distributions all configurations at
the Hamming distance $l$ from the reference sequence (selected with
$s_n=1, 1\le n\le N$) have one value of probability so the
probability of selecting the entire class of configurations is
$cp_l$. For symmetric fitness the mean fitness is calculated as in
Refs.~\cite{ba02,sh04c,ba98}:
\begin{eqnarray}
\label{e5}
\frac{R}{N}\equiv k= \max_{-1\le x\le1}U(x), \nonumber\\
U(x)= f(x)-1+\sqrt{1-x^2}.
\end{eqnarray}
The maximum point of Eq.(\ref{e5}) occurs at $x= x_c$. It follows
from Eq.(\ref{e3}) that $x_c$ can be interpreted as ``bulk
magnetization" in analogy with other models of statistical mechanics
\cite{ba97,bw01,ba98}:
 $$ x_c=\lim_{\beta\to\infty}\frac{ Tr \exp[-\beta
H]\sum_{k=1}^N\sigma^z_k}{N Tr \exp[-\beta H]}. $$
 Despite the lack of direct
biological meaning, we need to find $x_c$ to calculate the mean
fitness. For symmetric fitness function and permutation-invariant
original distribution there is a set of differential equations for
($N+1$) relative probabilities $p_l,0\le l\le N$ \cite{bw01}:
\begin{eqnarray}
\label{e6}
\frac{d{p_l}}{dt}= \nonumber \\
p_l [Nf(1-\frac{2l}{N})- N]+(N-l+1)p_{l-1}+(l+1)p_{l+1}.
\end{eqnarray}
Probability of finding all configurations at the Hamming distance
$l$ is $p_l/\sum_kp_k$. Mapping of the system of nonlinear equations
(1) onto the  system of linear equations (6) was calculated in
Refs.~[20-21]). In Eq.(6) we omit $p_{-1}$ and $p_{N+1}$ for $l=0$
and $l=N$, and set $\gamma_0=1$. In biological applications a
magnetization-like measure of surplus or surface magnetization can
be defined as
\begin{equation}
\label{e7} x_m=\frac{\sum_l(1-2l/N)p_l}{\sum p_l}.
\end{equation}
The main goal of this work is to calculate the dynamic of $x_m$ from
given initial distribution. Having the value of $x_c$ it is possible
to calculate the value of $x_m$ in steady state by solving:
\begin{equation}
\label{e8} f(x_m)=k.
\end{equation}
Various interpretations of bulk magnetization $x_c$ and surface
magnetization $x_m$ were analyzed in Ref.~\cite{ba98,bw01}. In next
sections we solve the model for the dynamics and determine explicit
role of $x_c$ for various sub-phases in dynamics.
\subsection{\bf HJE for Crow-Kimura model}
As in Ref.~\cite{sh07}, at a discrete $x=1-2l/N$ we use the ansatz
$p_l(t)\equiv p(x,t)\sim \exp[Nu(x,t)]$. Equation~(6) can be then
written as Hamilton-Jacobi equation for $u\equiv \ln p(x,t)/N$ (in
\cite{sh07} we gave an equation for individual probabilities in the
sequence):
\begin{equation}
\label{e9} \frac{\partial u}{\partial t}+H(u',x)=0,
\end{equation}
where $u'= \partial u/\partial x$,
\begin{equation}
\label{e10} -H(u',x)=f(x)-1+\frac{1+x}2e^{2u'}+\frac{1-x}2e^{-2u'},
\end{equation}
where the domain of $x$ is $ -1\le x\le 1$, and the initial
distribution is $u(x,0)=u_0(x)$.  Equation~(9) describes a class of
probabilities and the equation describing one sequence of
probabilities was given in Ref.~\cite{sh07}. In the limit of $t \to
\infty$ the asymptotic solution of Eq.~(9) is
\begin{eqnarray}
\label{e11} u(x,t;k) = kt+u_k(x),
\end{eqnarray}
where $u_k(x)$ can be calculated from Eq.(9) \cite{sh07} and the
mean fitness is $Nk$. Function $U(x)$ in Eq.(5) has a simple
physical interpretation as potential, i.e., the minimum of $-H(u,v)$
with respect to $v$ at a fixed $x$: $U(x)=min_v[- H(v,x)]$. It is
well-known from mechanics that motion is possible on an interval
when energy of the system is larger than potential $U(x)$ inside
this interval. In maximum-principle approach the largest eigenvalue
is identified with the mean fitness $k$. Similarly, $-k$ is the
maximal energy of the Hamiltonian $H(v,x)$ in Eq.~(10). A realistic
hypothesis would be to assume that the asymptotic solution
$u(x,t;k)$ is stable against perturbations only if $k$ is calculated
according to Eq.~(5). It is possible to obtain more results even
without solving the dynamics exactly. We know from physics that
motion in potential that has a single minimum is drastically
different from motion in potential with two or more minima.
Therefore, when in Fig.~1 function $U(x)$ changes from that depicted
by the continuous line to that presented by the dashed line but for
potential well $U(x)$ that has two maxima and two minima near $x=0$
we should anticipate phase transition.
\begin{figure}
\centerline{\includegraphics[width=0.6\columnwidth]{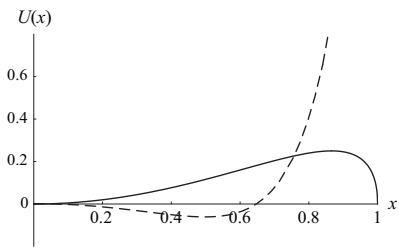}}
\caption{ Function $U(x) = f(x)+\sqrt{1-x^2}-1$ for $f(x)=x^2$
(solid curve) and for $f(x)=4\exp(-8+8x)$ (dashed curve). For the
latter there are two extrema where $U'(x)=0$: the maximum at
$0.9995$ (it is too high in is not shown in the graphics) and the
minimum at $0.497$. } \label{fig1}
\end{figure}
Here, we focus on the fitness $f(x)=cx^2/2$ [4] (the solid curve in
Fig.~1 corresponds to $c=2$). It results from Eq.~(5) that in this
case $U(x)$ has two extrema located on the interval $[-1;+1]$: the
minimum at $x=0$, and the maximum at $x=x_m$. To solve Eq.(\ref{e2})
subject to these initial data we use a standard procedure
\cite{mel98,eva02} by allowing to reduce the corresponding partial
differential equation to a system of ordinary differential
equations. Namely, consider the following set of equations:
\begin{equation}
\begin{aligned}
\label{e12}
\dot x =H_v(x,v)=-(1+x)\>e^{2v}+(1-x)\>e^{-2v},\\
\dot v = -H_x(x,v)=f'(x) + (e^{2v}-e^{-2v})/2, \\
\dot u = v\,H_v(x,v)-H(x,v)=v\dot x+q,
\end{aligned}
\end{equation}
subject to the following initial conditions: $x(0)=x_0$,
$v(0)=v_0(x_0)$, $u(0) = u_0(x_0)$. Here, $v=\partial u/\partial x$,
$v_0(x)= u_0'(x)$, and $q=\partial u/\partial t$. The corresponding
solution of Eq.~(\ref{e12}) in the $(x,t)$-space is called the {\it
characteristic} of Eq.~(\ref{e9}). Further, Eqs.~(\ref{e9}) and
(\ref{e12}) imply $\dot q=0$. Along the characteristic $x=x(t)$ and
variable $q$ is constant, so $q$ is selected to parameterize these
curves. Using the equation $q=f(x)-1+(1+x)/2e^{2v}+(1-x)/2e^{-2v}$,
we transform the first equation in Eq.(\ref{e12}) into
\begin{equation}
\label{e13} \dot x = \pm2\sqrt{[q+1-f(x)]^2 + x^2 - 1}.
\end{equation}
Having the solution of the characteristic system given by Eq.(12),
we can derive the solution of the original Eq.(\ref{e9})
\cite{mel98} by integrating the equation $\dot u = v\dot x+q$. For
biology applications it is important to know motions of distribution
maxima. For the purpose of finding these motions consider the
following initial distribution
\begin{equation}
\label{e14} u_0(x) = -a(x-x_0)^2.
\end{equation}
It is relatively easy to derive relaxation formulae for large values
of parameter $a$. We can calculate them directly from Eq.~(13),
using equation $q(x^*,t^*)=f(x^*)$ for the maximum point location
$x^*$. The maximum of the distribution moves along the branch of
Eq.(\ref{e13}) that preserves the sign of $x_0$. By integrating
Eq.(\ref{e13}) along the characteristic through the point
$(x^*,t^*)$ and assuming that $\dot x(t)$ does not change its sign,
we are getting
\begin{equation}
\label{e15} t^* =  \frac{\sign x_0}{2}\int\limits_{x^*}^{x_0}
\frac{d\xi}{\sqrt{(f(x^*)+1-f(\xi))^2+\xi^2-1}}.
\end{equation}
If at some point $x_1$ the characteristic $x(t)$ changes its
direction the point $x_1$ can be determined from the condition
\begin{equation}
\label{e16} [f(x^*)+1-f(x_1)]^2+x_1^2-1=0.
\end{equation}
In the latter case the integrals should be summed up over the
intervals $(x_0,x_1)$ and  $(x^*,x_1)$. This summation gives
\begin{equation}
\label{e17}
\begin{aligned}
t^* = \frac{\sign x_0}2&(\int\limits_{x_0}^{x_1}
\frac{d\xi}{\sqrt{(f(x^*)+1-f(\xi))^2+\xi^2-1}}\\+&
\int\limits_{x^*}^{x_1}
\frac{d\xi}{\sqrt{(f(x^*)+1-f(\xi))^2+\xi^2-1}}).
\end{aligned}
\end{equation}
Let $T_1$ be such that for $t\le T_1$ Eq.(\ref{e15}) holds, and for
$t>T_1$ Eq.(\ref{e17}) holds. At $T_1$ we have the condition
\begin{equation}
\label{e18} T_1=\frac{\sign x_0}{2}\int\limits_{X_1}^{x_0}
\frac{d\xi}{\sqrt{(f(X_1)+1-f(\xi))^2+\xi^2-1}},
\end{equation}
where $X_1$ is a root of $[f(X_1)+1-f(x_0)]^2+x_0^2-1=0$. For the
quadratic fitness $f(x)=cx^2/2$ with $c>0$ a selective phase exists
at $c>1$. Then, $x_m=1-\frac{1}{c}$ and $x_c=\sqrt{1-c^{-2}}$ [4].
When $t\rightarrow\infty$ the maximum converges to $x = x_m$. To
define the dynamics of the maximum at $-x_c\le x_0\le x_c$ we use
Eqs.(\ref{e15}) and (\ref{e17}), where
$$
x_1 = \sign
x_0\>\frac{\sqrt{c^2{x^*}^2+2(c-1)-2[(c-1)^2-c^2{x^*}^2]^{1/2}}}c
$$.
In the region where $x_c\le|x_0|\le1$ we use Eq.(15). To find $T_1$
in accordance with Eq.(\ref{e18}) we use
\begin{equation}
\label{e19} X_1 = \sign
x_0\sqrt{x_0^2-\frac{2[1-(1-x_0^2)^{1/2}]}c}.
\end{equation}
Figure~2 shows the evolution of the maximum for $c=2$ for $x_0 =
0,\,0.1,\,0.3,\,0.7,\,0.95$. These results demonstrate the excellent
agreement of analytic solutions given by Eqs.(15) and (17) with the
results of the numerical integration of Eq.(6). Note, Fig.~2 shows
that for $x_0<x_m$ the maximum moves initially away from the wild
configuration and returns to its neighborhood in later times. The
minimal $x^*(t)$ is just $X_1$.
\begin{figure}
\centerline{\includegraphics[width=0.6\columnwidth]{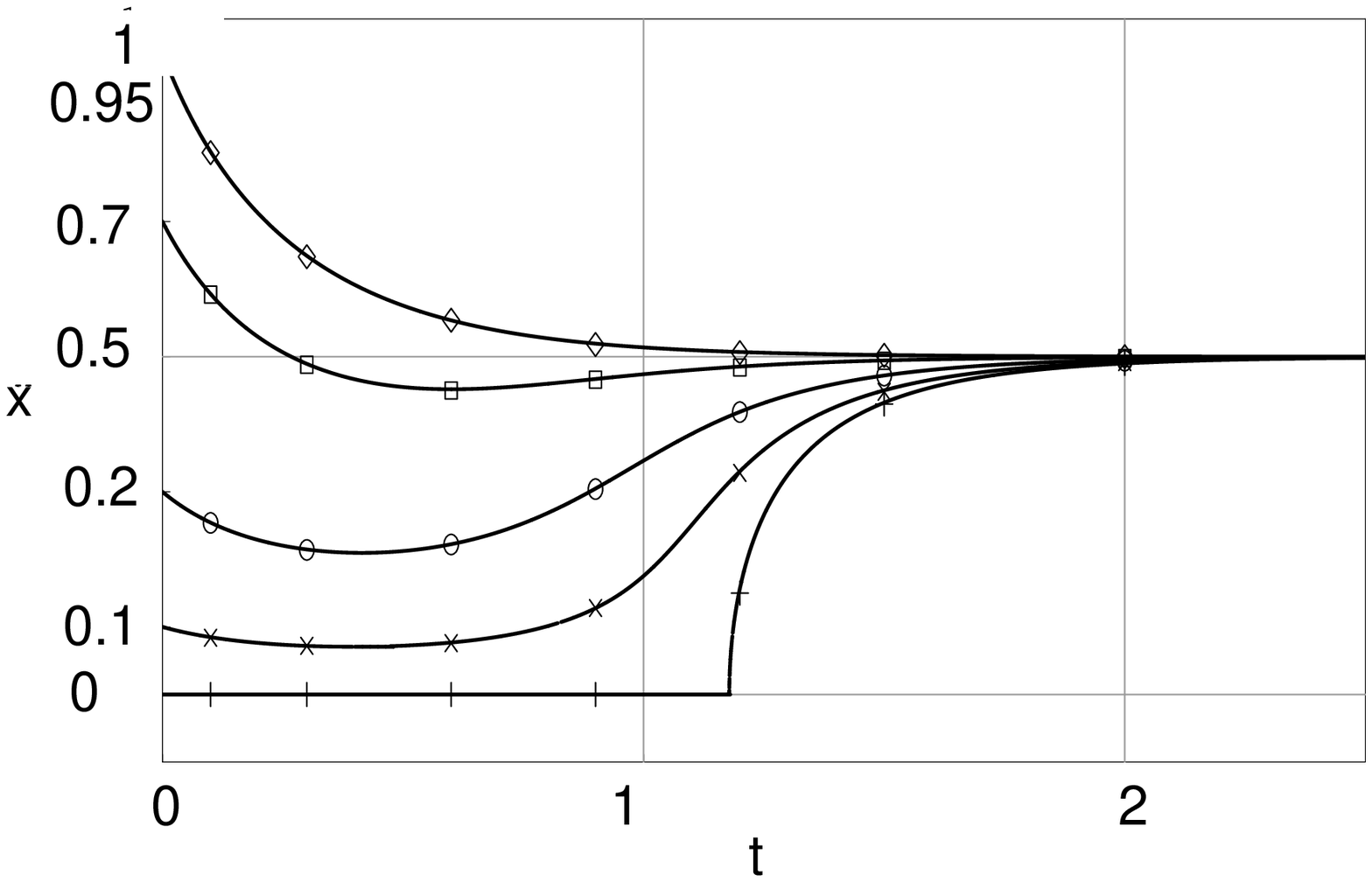}}
\caption{ The dynamics of the maximum point $x(t)$ for the
Crow-Kimura model ($f(x)=x^2$) for different initial values $x_0$ in
the distribution (14). The continuous curves are analytic results of
Eqs.(15) and (17). The symbols are the results of numerical
solutions of the Crow-Kimura model given by Eq.(6), where $N=1000$.
} \label{fig2}
\end{figure}
If $x^*(t)$ describes the position of maxima then
$v(x^*(t),t)=\frac{dv(x^*(t),t)}{dt}=0$ and Eqs.({\ref{e12}) give
\begin{eqnarray}
\label{20} \frac{dx^*(t)}{dt} = -2 x^*(t) -
\frac{f'(x^*(t))}{u_{xx}(x^*(t),t)},\quad x(0) = x_0,
\end{eqnarray}
where $u_{xx}(x,t) = {\partial v}/{\partial x}$. The motion of the
maximum of the distribution either towards the wild sequence or in
the opposite direction depends on the sign of
$f'(x^*(t))+2x^*(t)u''(x^*(t),t)$.
\subsection{\bf The flat original distribution}
When any of $2^N$ configurations is uniformly populated then the
initial condition for the entire probability class, having
probability $(^{\phantom{(1+x)}N!}_{(N(1+x)/2)!})\frac{1}{2^N}$,
yields
\begin{eqnarray}
\label{e21} u_0(x)= -\frac{1+x}{2} \ln\frac{1+x}{2}-\frac{1-x}{2}
\ln\frac{1-x}2.
\end{eqnarray}
Solution (21) has a peak at $x=0$. Let us calculate threshold-time
$T_2$ such that for $t\le T_2$ the population peak is in the class
of $x=0$. Assuming that at the moment $t^*$ the maximum is at point
$x^*$, we solve Eq.(13) for the characteristic with end-point
$(x^*,t^*)$ and, thus, take $q=f(x^*)$. The related characteristic
curve starts at the point $x(0)=x^*$, passes through the point
$(x_1,t^*/2)$ ($x_1$ is computed from Eq.(16)), turns, and finally
reaches the point $(x*,t*)$. Thus, Eq.(\ref{e16}) gives
\begin{equation}
\label{e22} t^* = {\sign x^*}\int\limits_{x^*}^{x_1}
\frac{d\xi}{\sqrt{(f(x^*)+1-f(\xi))^2+\xi^2-1}}.
\end{equation}
Now we take the limit as $x^*\to 0$ and find the threshold time
$T_2$. When $f(x)=cx^2/2$ and $c>1$ this time is
\begin{equation}
\label{e23} T_2 = {cos}^{-1}(\sqrt{1-1/c})/{\sqrt{c-1}}.
\end{equation}
\section{\bf The Eigen model}
As shown in Refs.~[5-6], for $2^N$ probabilities $P_i$ there is a
set of equations
\begin{equation}
\label{e24} \frac{dP_i}{d\tau}= \sum_{j=1}^{2^N}Q_{ij}r_j P_j-P_i[
\sum_{j=1}^{2^N}r_{j}P_j].
\end{equation}
Elements $Q_{ij}$ of the mutation matrix give the probabilities that
an offspring of configuration $j$ belongs to configuration $i$. In
this model mutations are quantified by
$Q_{ij}=q^{N-d(i,j)}(1-q)^{d(i,j)}$ and $\gamma=N(1-q)$, where
$\exp[-\gamma]\equiv q^N$ is the probability of having exact copy,
$r_j=f(1-2l/N)$ is the fitness, and $l$ is the Hamming distance of
the $j$th  configuration from the reference configuration. The
Hamming distance between configurations $i$ and $j$ (that have spins
spins $s^i_n$ and $s^i_n$, respectively) is
$d(i,j)=\sum_n(1-s^i_ns^j_n)$. Considering again the ($N + 1$)
Hamming-class probabilities $p_l$ for $p_l\equiv \exp[Nu(x,t)] and
x=1-2l/N$, Eq.(24) of Ref.~\cite{sh07} has been mapped onto the
following equation
\begin{eqnarray}
\label{e25} {\frac {\partial u}{\partial t}} =f(x) e^{\gamma [
\mathrm{ch} (2u')+x \mathrm{sh} (2u') -1]},
\end{eqnarray}
where $\tau=tN$. Asymptotic solutions $u(x,t;k)=kt+u_k(x)$ ($k$ is a
mean fitness \cite{sh06}) in the limit of $t\to \infty$ are as
follows
\begin{eqnarray}
\label{e26} k= \max_{-1\le x\le1}U(x),\quad U(x)=f(x)\exp(\gamma
[-1+\sqrt{1-x^2}]) ,
\end{eqnarray}
where $x_c$ and $x_m$ are obtained from
\begin{equation}
\label{e27}
\begin{aligned}
U'(x_c)=0,\quad f(x_m)=f(x_c)\exp(-\gamma [1-\sqrt{1-x_c^2}]).
\end{aligned}
\end{equation}
When $x_c<|x_0|<1$ then for initial distribution given by Eq.(14)
with $a >> 1$ the position of the maximum $(t*,x*)$ is
\begin{equation}
\label{e28} t^*=\frac{\sign
x_0}2\int\limits_{x^*}^{x_0}\,\frac{d\xi}
{f(x)\,\sqrt{\left(\ln\frac{f(x)}{f(\xi)}+\gamma\right)^2-{\gamma}^{2}(1-\xi^2)}
}.
\end{equation}
For all other cases the solution is
\begin{equation}
\label{e29}
\begin{aligned}
t^*=\frac{\sign
x_0}2&\Big(\int\limits_{x_0}^{x_1}\,\frac{d\xi}{f(x^*)\,
\sqrt{\left(\ln\frac{f(x^*)}{f(\xi)}+\gamma\right)^2-{\gamma}^{2}(1-\xi^2)}}+\\
{}+&\int\limits_{x^*}^{x_1}\,\frac{d\xi}{f(x^*)\,
\sqrt{\left(\ln\frac{f(x^*)}{f(\xi)}+\gamma\right)^2-{\gamma}^{2}(1-\xi^2)}}\Big
),
\end{aligned}
\end{equation}
where $x_1$ can be calculated from the condition
\begin{eqnarray}
\label{e30}
\left(\ln\frac{f(x^*)}{f(x_1)}+\gamma\right)^{\!2}-{\gamma}^{2}(1-x_1^2)=0.
\end{eqnarray}
Finally, for relaxation from the flat distribution we get:
\begin{eqnarray}
\label{e31} t^*=\sign{x^*}\int\limits_{x^*}^{x_1}\,\frac{d\xi}
{f(x^*)\,\sqrt{\left(\ln\frac{f(x^*)}{f(\xi)}+\gamma\right)^{\!2}-{\gamma}^{2}(1
-\xi^2)}},
\end{eqnarray}
\begin{figure}
\large \unitlength=0.1in
\begin{picture}(42,12)
\put(-2.2,-13.5){\includegraphics{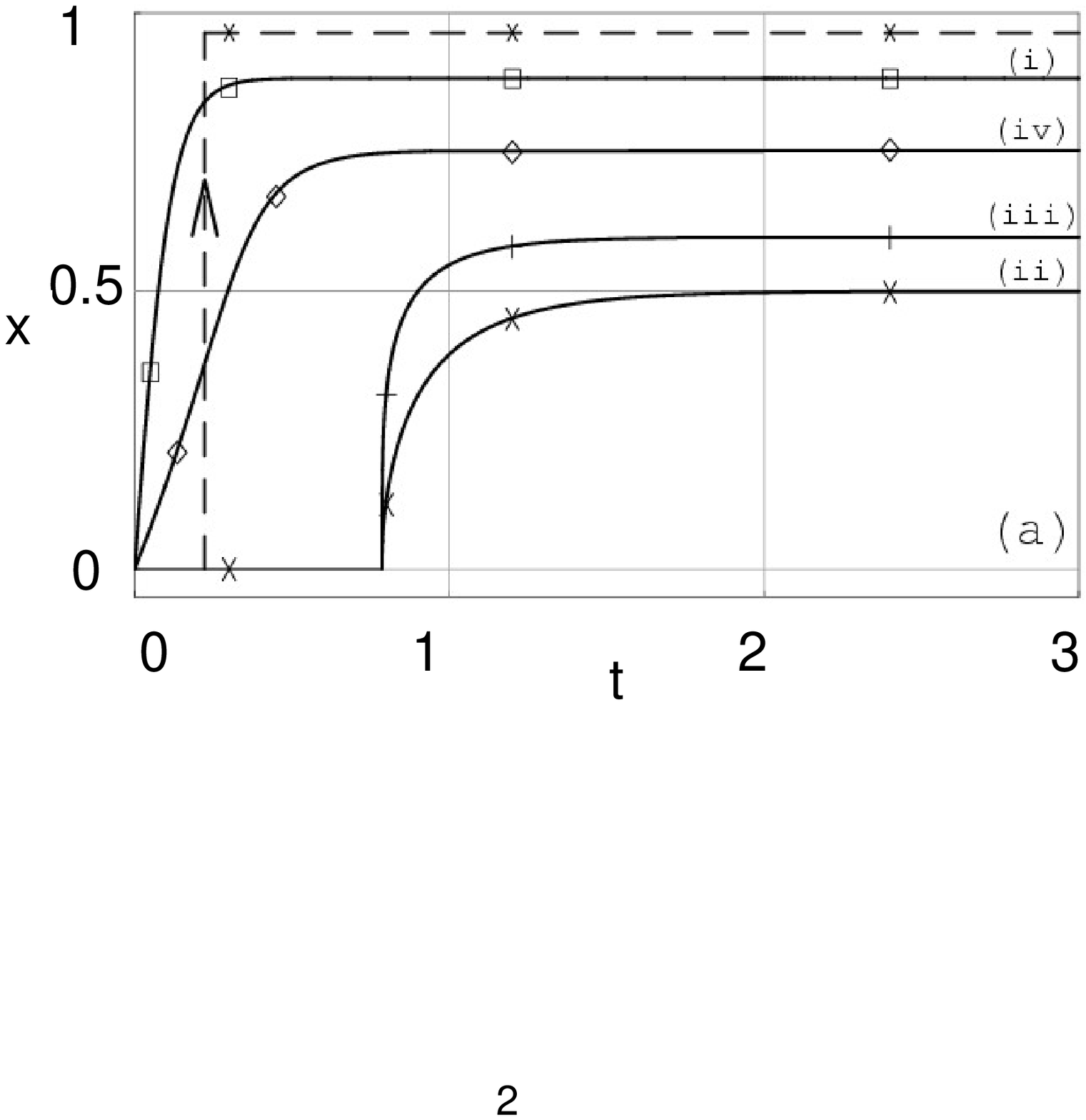}}
\put(16.5,-15){\includegraphics{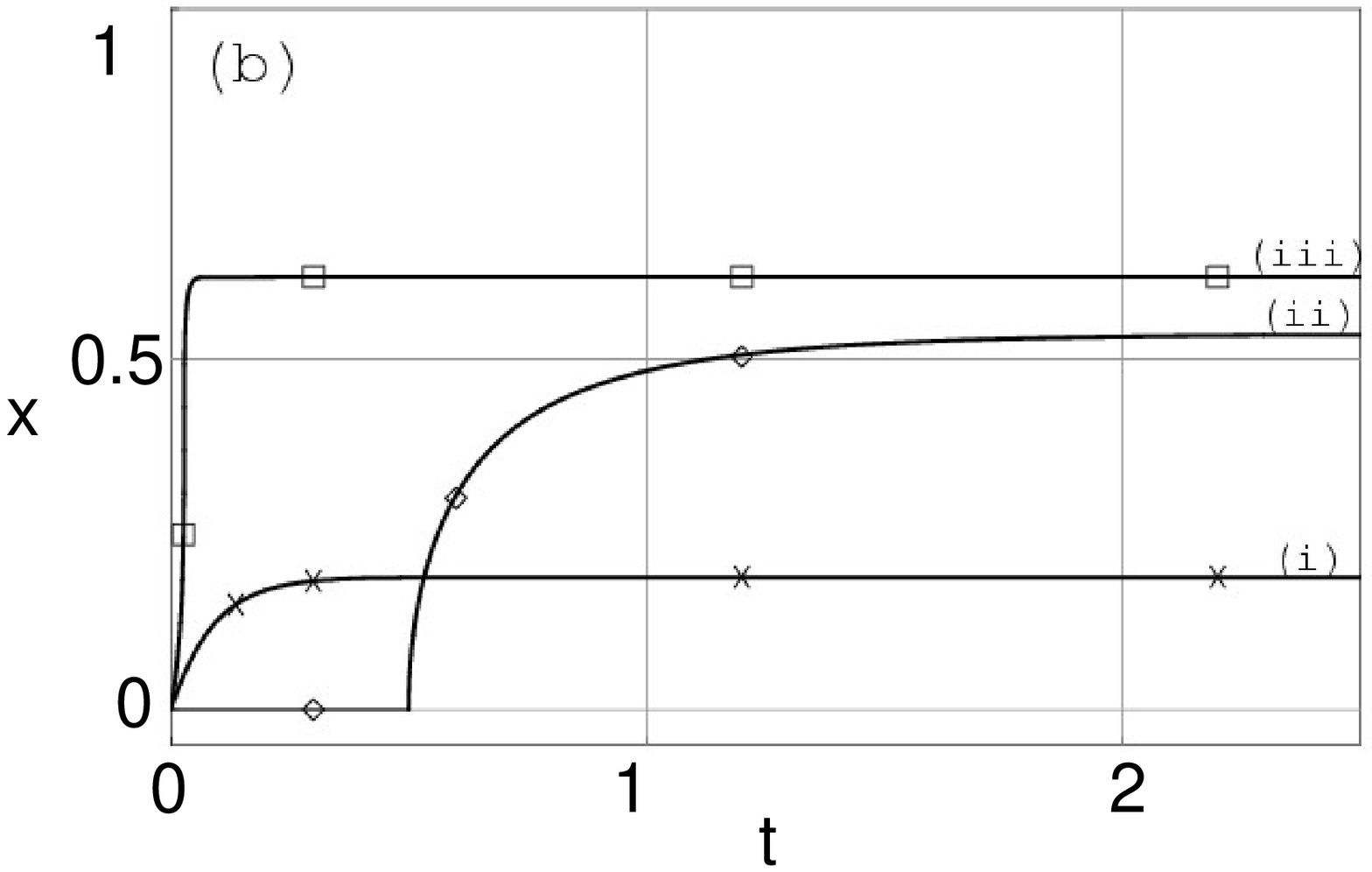}}
\end{picture}
\caption{Dynamics of maximum density points $x^*(t^*)$ for the flat
initial distribution given by Eq.(14): (a) Crow-Kimura model where
(i) $f(x)=8x$, (ii) $f(x)=x^2$, (iii) $f(x)=x^2+0.2x^4$, (iv)
$f(x)=4\exp(x-1)$, and $f(x)=4\exp(-8[1-x])$ (dashed line); (b)
Eigen model where $\gamma=2$ and (i) $f(x)=2(x+1)$, (ii) $f(x)=x^2$,
and (iii) $f(x)=\exp(4x)$. Continuous curves are the analytical
results. The symbols are the solutions of numerical integration.}
\end{figure}
\begin{figure}
\centerline{\includegraphics[width=0.6\columnwidth]{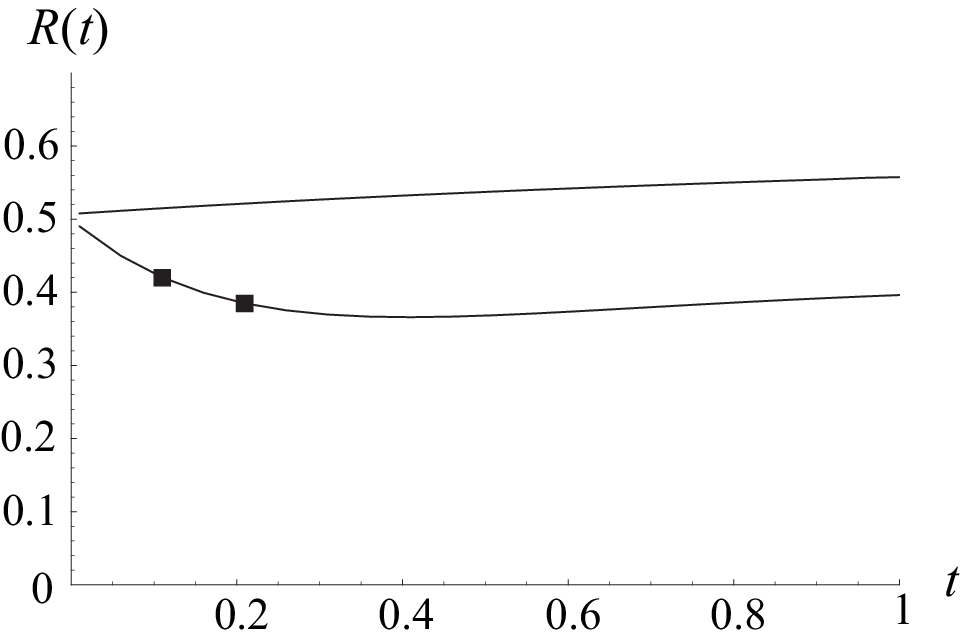}}
\caption{ The dynamics of the mean fitness $R(t)$ for the
Crow-Kimura model ($f(x)=x$) for different initial values $x_0=0.5$
in the distribution (14). The symbols are the results of numerical
solutions of the Crow-Kimura model given by Eq.(6), where $N=1000$.
The upper line is an approximate result by diffusion method, the
lower line is our exact result. } \label{fig5}
\end{figure}
\begin{figure}
\centerline{\includegraphics[width=0.6\columnwidth]{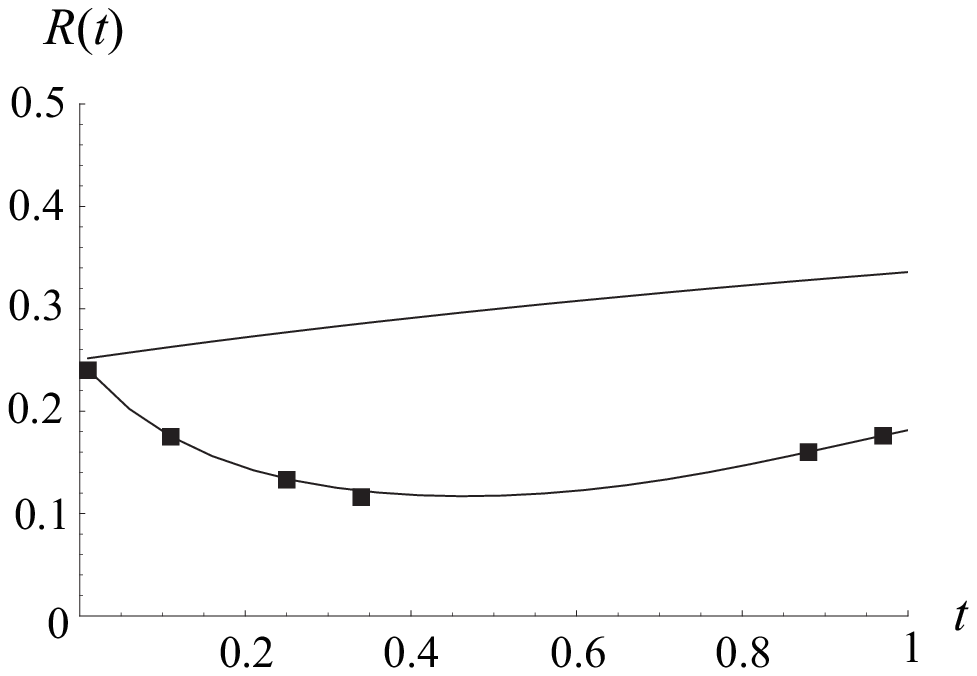}}
\caption{ The dynamics of the mean fitness $R(t)$ for the
Crow-Kimura model ($f(x)=x^2$) for different initial values
$x_0=0.5$ in the distribution (14). The symbols are the results of
numerical solutions of the Crow-Kimura model given by Eq.(6), where
$N=1000$. The upper line is an approximate result by diffusion
method, the lower line is our exact result. } \label{fig6}
\end{figure}
\begin{center}
\begin{table}[t]
\begin{tabular}{|c|c|c|c|c|c|c|c|}
\hline $c$          & 1.1   & 1.2   & 1.3   & 1.4  & 1.5     & 1.6 \\
\hline $T_2$        &2.397 &1.791  & 1.466 &1.252 & 1.098   &0.980  \\
\hline $t_2$        &3.998 & 2.572 &1.953  & 1.591 & 1.351 & 1.177 \\
\hline
\end{tabular}
\caption{ Comparison of $t_2$, the result of \cite{ba98} for the
threshold time period in case of initially flat distribution, with
$T_2$, our exact result by Eq.(23) for Crow-Kimura model with
$f(x)=cx^2/2$. }
\end{table}
\end{center}
\section{\bf Discussion}
We have considered discrete-error classes in continuum
approximation, replacing the system of equations for molecular
evolution by a single Hamilton-Jacobi equation. Dynamics have been
obtained by solving this equation. This method is qualitatively
similar to semi-classical methods, well-known in quantum mechanics.
Our approach has an accuracy of $1/N$, where $N$ is genome length.
There is straightforward connection between our current method and
methods that utilize statistical-physics analogies with Ising spins.
Specifically, two different sub-phases that have been determined
with our method describe two different relaxation regimes (i.e.,
Eqs.(15) and (17) for Crow-Kimura's model; and, Eqs.(28-30) for
Eigen's model). These two relaxation regimes correspond exactly to
two different magnetization values as discussed in
Refs.~\cite{bw01,ba02,ba07}. Singularities $x_c$ in relaxation
periods correspond to bulk magnetization. Initially, when the entire
virus population is in one genetic configuration that is closer to
the wild configuration than the sequences with the same value of
$x_c$, the maximum in the population distribution moves to the
steady state $x_m$. This is in analogy with surface magnetization.
On the other hand, when the initial configuration is far away from
$x_c$, the maximum of the distribution moves away from the wild
configuration in the initial phase and moves towards $x_m$ in a
later phase. The single minimum at $x=0$ of the evolution potential
$U(x)$ (i.e., Eq.(5) for Kimura's model, and Eq.(26) for Eigen's
model) gives smooth dynamics (see Eqs.(15), (17), Fig.~2, and
Eqs.(28-30)). Equations (22) and (31) give the evolution from the
original flat distribution in the Crow-Kimura and the Eigen models,
respectively. These results are presented in Fig.~3 for several
choices of fitness function. Analytical dynamics of maximum-density
points $x^*(t^*)$ is in excellent agreement with numerical solutions
for the original formulation of these models. The second phase of
the dynamics with a jump in the position of $x^*(t)$ (seen as the
dashed line in Fig.~3a) is related to the presence a potential well
(indicated by the dashed line in Fig.~1). Preliminary numerical
studies of similar problems indicate the existence of a similar
phase with a jump that does not require a potential well but a steep
potential. The evolution dynamics is a highly non-trivial
phenomenon. As we demonstrated in this work, even for monotonic and
smooth fitness landscapes it is possible to have discontinuous
dynamics in analogy with the punctuated evolution of
Ref.~\cite{dr01} or the shock waves of Ref.~\cite{ca05}). Such
discontinuous dynamics for smooth fitness function has been also
found in Ref.~\cite{ba01}, where the dynamic of the evolution model
was investigated numerically for four-valued spins. In the current
article we suggest the analytical method to investigate
discontinuous evolution for a general fitness case. In
Ref.~\cite{ba98} an analytic approximation that would be accurate
for large $c$ have been suggested for the dynamic of Crow-Kimura
model. In Table 1 we compare our exact result for $T_2$ obtained
from Eq.(23) with the corresponding expression derived by the method
of Ref.~\cite{ba98} (by setting $\lambda=1$ in Eqs.(4) and (65) of
Ref.~\cite{ba98}). Our method gives the full distribution, while the
method of Ref.~\cite{ba98} gives the position of the distribution
maximum. In summary, we considered HJE to obtain exact dynamics and
used Hamiltonian mechanics for qualitative analysis of evolution
models. Our results are valid for any analytic fitness function. The
diffusion method of Refs.~[10-13] is valid only near the maximum of
distribution or for the case of weak selection, and yields
inaccurate results when applied for long relaxation periods or for
calculating mean fitness. This yields the error greater than $50\%$
after $t=0.2$ (see Figs.~4-5). The HJE approach is self-consistent,
with no need to use genome length (which is in contrast to
Refs.~[10-13]), and gives the dynamic with the $1/N$-accuracy.

We thank M.\,W.~Deem, A.~Kolakowska, A.~Melikyan, L.~Peliti,
S.~Nazarian, and D.~Waxman for discussions. D.~B.~Saakian thanks the
Volkswagenstiftung grant ``Quantum Thermodynamics", U.S. Civil
Research Development Foundation ARP2-2647-Ye-05, National Center for
Theoretical Sciences in Taiwan and Academia Sinica (Taiwan), Grant
No. AS-95-TP-A07.

\end{document}